\documentclass[preprint,showpacs,showkeys,preprintnumbers,amsmath,amssymb]{revtex4}
 \usepackage{dcolumn}% Align table columns on decimal point
 \usepackage{bm}% bold math

\begin{document}

 \title{ Quasinormal modes and entropy spectrum
 of three dimensional G\"{o}del black hole }

 \author{ Ran Li }

 \thanks{Electronic mail: liran.gm.1983@gmail.com}

 \affiliation{Department of Physics,
 Henan Normal University, Xinxiang 453007, China}

 \begin{abstract}

 We have studied perturbations of scalar and
 spinor field in the background of three dimensional
 G\"{o}del black hole. The wave equations 
 are shown to be exactly solvable in terms of
 hypergeometric functions. The quasinormal modes 
 are analytically calculated
 by imposing the Dirichlet boundary condition
 at spatial infinity,
 which are shown to be of the same form in both cases.
 By considering the physical interpretation of quasinormal modes,
 we obtain the consistent transition frequencies
 from the quasinormal modes of scalar and spinor field.
 As an application of quasinormal modes,
 we have also investigated the area and entropy
 quantization of three dimensional
 G\"{o}del black hole.
 By choosing the conserved mass of
 G\"{o}del black hole properly, the entropy spectrum
 are shown to be equally-spaced.

 \end{abstract}



 \maketitle
 \newpage

 \section{Introduction}

 The studying of matter field perturbation in black hole
 background can provide deep insight into
 quantum properties of black hole. By using
 the techniques of quantum field theory in curved spacetime,
 Hawking \cite{hawking} made the intriguing observation that
 black hole can radiate thermally like a blackbody at
 the temperature propotional to the surface gravity
 of the event horizon. This discovery confirms
 the Bekenstein's conjecture \cite{ben} that black hole
 have a thermodynamics entropy propotional
 to the area of horizon.

 Quasinormal mode is an interesting topic
 in black hole perturbation theory.
 It can be regarded as
 the resonance of a black hole under a small
 external field perturbation.
 Quasinormal modes can be computed
 by solving the perturbation field equation in a fixed
 background spacetime.
 Quasinormal modes are damped modes with complex frequencies,
 which depend only on the parameters of black hole.
 The studying of quasinormal modes has a long
 history, and there has been extensive work done to compute
 quasinormal modes in various black hole backgrounds.
 For a recent review, one can refer to \cite{berti,qnmreview}.

 Inspired by the AdS/CFT correspondence \cite{meldcena,gubser,witten},
 quasinormal modes of asymptotically AdS black holes
 are related to the retarded
 Green's functions of the dual thermal conformal field theories.
 It was established \cite{horowitz} that
 the relaxation time of a thermal state of the boundary thermal
 field theory is proportional to the inverse of the imaginary part of
 quasinormal modes of the dual gravity background.
 Besides their theoretical importance, quasinormal modes also have their
 astronomical applications. One can
 get a deeper understanding of the black holes in nature
 by detecting signals of quasinormal modes using
 the gravitational wave detectors \cite{abbott}.

 In recent years, it was proposed that quasinormal
 modes can also be used to quantize the area and the entropy
 of a black hole. Bekenstein \cite{bekenstein} is the first to
 propose that the area of a black hole should be quantized.
 He showed that the black hole horizon area is an adiabatic
 invariant, which leads to a discrete spectrum
 under a suitable quantization process.
 Hod's proposal of the relationship between quasinormal modes
 and area quantization of a black hole suggests that
 the real part of the asymptotic
 quasinormal modes can be regarded
 as a transition frequency in the
 semiclassical limit \cite{Hod}.
 Latter, Kunstatter \cite{kunstatter}
 suggested that the horizon area of a black hole
 can be quantized by using an adiabatic invariant $I$.
 For a system with energy $E$ and vibrational frequency
 $\omega(E)$,  the adiabatic invariant $I=\int dE/\omega(E)$
 is quantized by identifying the real part of the
 quasinormal modes as the transition frequency
 and using Bohr-Sommerfeld quantization condition.
 More recently, Maggiore \cite{maggiore}  proposed that,
 for a highly damped modes, the imaginary part
 of quasinormal modes should be taken as the transition frequency.
 Because of these developments, there have been
 many works to compute quasinormal modes
 and area spectra in various types of black holes \cite{manywork}.

 Three dimensional G\"{o}del spacetime \cite{banados} is an
 exact solution to Einstein-Maxwell
 theory in 2+1 dimensions with
 a negative cosmological constant and a
 Chern-Simons term. This theory can be viewed
 as a lower dimensional toy model for
 the bosonic part of five dimensional supergravity theory,
 since the field content and
 the couplings of both theories are similar.
 Three dimensional G\"{o}del black holes
 display the same peculiar
 properties as their higher dimensional
 counterparts\cite{banados}.
 The rotating black hole solutions on the G\"{o}del background
 in the context of five-dimensional supergravity theory
 have attracted a lot of attention \cite{godelsolution}.
 More recently, the quasinormal modes and stability of
 five-dimensional rotating G\"{o}del black holes are
 investigated by R. A. Konoplya et.al. in \cite{godelqnm}.
 In this paper, we will
 study the perturbations of scalar field and
 spinor field in the background of three dimensional
 G\"{o}del black hole.

 Most calculations of quasinormal modes
 are numerical due to the difficulties in solving the differential
 equations. However, in three dimensions, there have been
 several papers where quasinormal modes are computed
 analytically. The well-known BTZ black hole
 has been studied with exact results \cite{BTZ}.
 Recent studies show that the quasinormal modes
 of warped AdS black holes can also be
 analytically calculated \cite{warpedads,ranli}.
 We will show that the wave equations of scalar field and spinor
 field in three dimensional G\"{o}del black hole can
 also be exactly solved in terms of
 hypergeometric functions.
 The quasinormal modes of
 scalar field and spinor field are analytically calculated
 by imposing the Dirichlet boundary condition
 at the spatial infinity,
 which are shown to be of the same form in both cases.
 By considering the physical interpretation of quasinormal modes,
 we obtain the consistent transition frequencies
 from the quasinormal modes of scalar field and spinor field.
 As an application of quasinormal modes,
 we use the proposal of area quantization of rotating
 black hole in \cite{cqg1,cqg2} to investigate area and entropy
 quantization of three dimensional
 G\"{o}del black hole. It is shown that, when the conserved mass
 of three dimensional G\"{o}del black hole is taken as the parameter
 $\nu$, the area and entropy can be quantized and
 the spectra are equally-spaced.

 This paper is arranged as following.
 In Sec. II, we give a brief review of
 three dimensional G\"{o}del black hole.
 In Sec. III and IV, we consider
 the perturbations of scalar field
 and spinor field in the background of three
 dimensional G\"{o}del black hole and calculate
 the corresponding quasinormal modes
 analytically by solving the equation of motion.
 In Sec. V, using the computed quasinormal modes,
 area and entropy spectrum of
 three dimensional G\"{o}del black hole are presented.
 The last section is devoted to summary and conclusion.

 \section{Three Dimensional G\"{o}del black hole}

 In this section, we will firstly give a brief review
 of the geometric and thermodynamic properties of
 three dimensional G\"{o}del black hole.
 Three dimensional G\"{o}del spacetime is an
 exact solution to Einstein-Maxwell
 theory in 2+1 dimensions with
 a negative cosmological constant and a
 Chern-Simons term. The action is given by
 \begin{eqnarray}
 I=\frac{1}{16\pi G}\int d^3 x\left[
 \sqrt{-g}\left( R+\frac{2}{l^2}
 -\frac{1}{4}F_{\mu\nu}F^{\mu\nu}\right)
 -\frac{\alpha}{2}\epsilon^{\mu\nu\rho}
 A_{\mu}F_{\nu\rho}\right]\;,
 \end{eqnarray}
 where $G$ is the three dimensional gravitational constant,
 the parameter $l$ is related to the cosmological
 constant $\Lambda$ by $\Lambda=-1/l^2$ and 
 $\alpha$ is the Chern-Simons coupling constant.

 It has been shown that the equations of motion
 derived from this action admit the black hole
 solution \cite{banados}, where the metric and 
 the non-vanishing component of
 gauge potential $A_{\mu}$ are given by
 \begin{eqnarray}
 ds^2&=&dt^2-4\alpha r dt d\varphi
 +\left[ 8G\nu r -(1-\alpha^2 l^2)
 \frac{2r^2}{l^2}-\frac{4GJ}{\alpha} \right]
 d\varphi^2 \nonumber\\
 &&+\left[ (1+\alpha^2 l^2)\frac{2r^2}{l^2}
 -8G\nu r+\frac{4GJ}{\alpha}
 \right]^{-1}dr^2\;,
 \\
 A_{\varphi}&=&-\frac{4GQ}{\alpha}
 +\sqrt{1-\alpha^2 l^2}\frac{2r}{l}\;.
 \end{eqnarray}
 The parameters $\nu$, $J$ and $Q$ are 
 integral constants, which may be related to 
 the mass, angular momentum and charge of black hole.  

 Note that because of the presence of
 a nontrivial gauge field, the
 asymptotic geometry of three dimensional G\"{o}del black hole
 does not behave as either de Sitter
 or anti-de Sitter.
 It is also obvious that
 it is not asymptotic to
 the warped AdS spacetime.
 The asymptotic symmetry algebra of this
 spacetime has been studied in \cite{compere},
 which turns out to be
 the semi-direct sum of the diffeomorphisms
 on the circle with two loop algebras.
 The covariant Poisson bracket of
 the conserved charges associated with
 the generators of asymptotic symmetry group
 is shown to be centrally extended
 to the semi-direct sum of a Virasoro algebra
 and two affine algebras. The aspect of
 holographic dual between
 this black hole and conformal field theory
 should be further studied, which is not considered
 in the present paper.

 The black hole has two horizons, i.e.
 the inner and the outer event horizons
 $r_{\pm}$, which are determined by the equation
 \begin{eqnarray}
 (1+\alpha^2 l^2)\frac{2r^2}{l^2}
 -8G\nu r+\frac{4GJ}{\alpha}=0\;.
 \end{eqnarray}
 The solutions give the locations of event horizons
 \begin{eqnarray}
 r_{\pm}=\frac{l^2}{1+\alpha^2 l^2}
 \left[ 2G\nu\pm\sqrt{4G^2\nu^2-
 \frac{2GJ}{\alpha}\frac{(1+\alpha^2 l^2)}{l^2}}
  \right]\;.
 \end{eqnarray}
 The outer and the inner event horizons are the coordinate
 singularities of the metric, which can be eliminated
 by a proper coordinates transformation.

 Now, Let us discuss the thermodynamics of the black hole.
 The Hawking temperature $T_H$ is computed as
 \begin{eqnarray}
 T_H&=&\frac{(1+\alpha^2 l^2)}{4\pi\alpha l^2}
 \frac{(r_+-r_-)}{r_+}\;.
 \end{eqnarray}
 The angular velocity at the outer event horizon
 of the black hole is given by
 \begin{eqnarray}
 \Omega_H=\frac{1}{2\alpha r_+}\;.
 \end{eqnarray}
 The entropy of the black hole can be calculated
 by using the Wald's formalism, which is the quarter
 of area of the outer horizon for Einstein gravity
 \begin{eqnarray}
 S=\frac{A}{4G}=\frac{\pi\alpha r_+}{G}\;.
 \end{eqnarray}

 Here, we present an intuitive derivation of
 conserved charges of the black hole.
 By differentiating the expression of the outer
 event horizon $r_+$,
 one can deduce the following relationship
 \begin{eqnarray}
 d\nu=TdS+\Omega_H dJ\;.
 \end{eqnarray}
 If one identifies the parameter $\nu$ as the mass $M$
 and the parameter $J$ as the
 angular momentum of the black hole,
 this differential relationship can be
 treated as the first law of black hole
 thermodynamics. It has been shown in \cite{banados} that,
 via the rigorous definition of conserved charges
 and tensor calculation, the parameter $\nu$ is
 the conserved quantity associated to the
 killing vector $\partial_t$. Here, we just present
 a simple and intuitive derivation of black hole mass.
 However, it is observed in \cite{compere} that,
 under the change of coordinates $r\rightarrow-r$,
 $\phi\rightarrow-\phi$, the solutions with
 the parameters $(\nu, J, Q)$ can be changed to
 the solutions with the parameters $(-\nu, J, -Q)$.
 So the conserved quantity $\nu$ does not provide
 a satisfactory definition of black hole mass.
 They found the definition of black hole mass
 as $\mu=2G\nu^2$ can properly recover this shortcoming.
 As we will done in the Sec.V,
 the conserved quantity $\nu$ has been taken as the energy
 of black hole in the process of entropy quantization.
 The result shows this identification is effective and
 reasonable, at least, in handling with
 the quantization of black hole entropy.

 One should also note that
 the electrostatic potential and charge does not appear
 in the differential form of thermodynamics first law.
 In fact, one can impose a background electrostatic potential
 to make the total electrostatic potential vanish.
 Especially, the electric parameter $Q$ is an integral constant
 and dose not appear in the metric function.
 So, when calculating the quasinormal modes of
 perturbation fields, we will not consider
 the coupling between the gauge field $A_\mu$ and the matter fields.

 \section{quasinormal modes of scalar field }

 In this section, we will analytically calculate quasinormal modes of scalar
 field perturbation in the background of three dimensional G\"{o}del black hole.
 We consider the equation of motion for scalar field perturbation,
 which is given by the Klein-Gordon equation
 \begin{eqnarray}
 \frac{1}{\sqrt{-g}}\partial_\mu\left(
 \sqrt{-g}g^{\mu\nu}\partial_{\nu}\Phi
 \right)+\mu^2 \Phi=0,
 \end{eqnarray}
 with $\mu$ being the mass of scalar field.

 According to the symmetry of background spacetime,
 i.e. the existence of two killing vectors
 $\partial_t$ and $\partial_\varphi$,
 we can expand the scalar
 field $\Phi(t, r, \varphi)$
 as following
 \begin{eqnarray}
 \Phi(t, r, \varphi)=e^{-i\omega t+im\varphi}R(r)\;.
 \end{eqnarray}
 Then, by substituting this expression for the
 scalar field perturbation
 into Eq.(10),
 one can get the radial wave equation
 after some algebra
 \begin{eqnarray}
 \Delta\frac{d}{dr}\left(\Delta\frac{d}{dr} R(r)\right)
 +\left(\omega^2(4\alpha^2 r^2-\Delta)-4\omega m \alpha r
 +m^2-\mu^2\Delta
 \right)R(r)=0,
 \end{eqnarray}
 where, for latter convenience, we have introduced the function
 $\Delta(r)=\lambda(r-r_+)(r-r_-)$ with
 $\lambda=2(1+\alpha^2 l^2)/l^2$.
 This equation can be analytically solved by the
 hypergeometric function,
 which provides us with an exact calculation of
 quasinormal modes of scalar field perturbation.

 In order to solve the radial wave equation,
 it is convenient to introduce the variable $z$ as
 \begin{eqnarray}
 z=\frac{r-r_+}{r-r_-}\;.
 \end{eqnarray}
 Then, the radial wave equation can be rewritten in the form
 of hypergeometric equation
 \begin{eqnarray}
 z(1-z)\frac{d^2 R(z)}{dz^2}+
 (1-z)\frac{dR(z)}{dz}+\left(
 \frac{A}{z}+B+\frac{C}{1-z}\right)R(z)=0\;,
 \end{eqnarray}
 where the parameters $A$, $B$ and $C$ are given by
 \begin{eqnarray}
 A&=&\frac{\left(2\alpha r_+\omega-m\right)^2}{\lambda^2(r_+-r_-)^2}
 \;,\nonumber\\
 B&=&-\frac{\left(2\alpha r_-\omega-m\right)^2}{\lambda^2(r_+-r_-)^2}
 \;,\nonumber\\
 C&=&\frac{4\alpha^2\omega^2}{\lambda^2}-\frac{\omega^2+\mu^2}{\lambda}\;.
 \end{eqnarray}

 In terms of the definition of quasinormal modes,
 the solution of perturbation field
 must be purely ingoing near the horizon of black hole.
 So we are just interested in the solution
 with the ingoing boundary condition
 at the horizon.
 The solution of radial wave equation with the ingoing
 boundary condition is given explicitly by the hypergeometric
 function
 \begin{eqnarray}
 R(z)=z^{\alpha_s}(1-z)^{\beta_s}F(a_s,b_s,c_s,z)\;,
 \end{eqnarray}
 where
 \begin{eqnarray}
 \alpha_s=-i\sqrt{A}\;,\;\;\;
 \beta_s=\frac{1}{2}-\sqrt{\frac{1}{4}-C}\;,
 \end{eqnarray}
 and
 \begin{eqnarray}
 c_s&=&2\alpha_s+1\;,\nonumber\\
 a_s&=&\alpha_s+\beta_s+i\sqrt{-B}\;,\nonumber\\
 b_s&=&\alpha_s+\beta_s-i\sqrt{-B}\;.
 \end{eqnarray}

 So, we have shown that
 the equation of motion
 for scalar field perturbation
 in the background of three dimensional
 G\"{o}del black hole
 can be exactly solved in terms
 of hypergeometric function
 after the partial wave decomposition.
 Now, we will analyse the asymptotic
 properties of the solution at spatial
 infinity and calculate
 the corresponding quasinormal modes
 by imposing the Dirichlet boundary
 condition at spatial infinity.

 By using the following transformation property
 of hypergeometric function
 \begin{eqnarray}
 &&F(a,b,c;z)=\frac{\Gamma(c)\Gamma(c-a-b)}{\Gamma(c-a)\Gamma(c-b)}
 F(a,b,a+b-c+1;1-z)\nonumber\\
 &&\;\;\;\;+(1-z)^{c-a-b}
 \frac{\Gamma(c)\Gamma(a+b-c)}{\Gamma(a)\Gamma(b)}
 F(c-a,c-b,c-a-b+1;1-z)\;,
 \end{eqnarray}
 one can find the leading asymptotic behaviour
 of the solution $R(z)$ at the spatial infinity
 $(\textrm{i.e.}, z\rightarrow 1)$
 \begin{eqnarray}
 R(z)\simeq z^{\alpha_s}(1-z)^{\beta_s}\frac{\Gamma(c_s)\Gamma(c_s-a_s-b_s)}
 {\Gamma(c_s-a_s)\Gamma(c_s-b_s)}\;.
 \end{eqnarray}

 Next, in order to find the quasinormal modes,
 one has to impose the boundary condition at the asymptotic
 infinity. The condition that the flux
 vanishes at asymptotic infinity
 is just a perfect one. Here,
 we will use the equivalent
 Dirichlet condition that the scalar field is
 vanishing at asymptotic infinity
 to obtain the quasinormal modes.
 Imposing the vanishing Dirichlet boundary
 condition at spatial infinity leads to the following
 relation
 \begin{eqnarray}
 c_s-a_s=-n\;,\;\;\textrm{or}\;\;c_s-b_s=-n\;,
 \end{eqnarray}
 which give the following two equations of quasinormal modes for scalar
 field perturbation
 \begin{eqnarray}
 &&\omega-\frac{m}{\alpha(r_++r_-)}+i\frac{\lambda(r_+-r_-)}{2\alpha(r_++r_-)}(n+h_s)
 =0,\nonumber\\
 &&\omega+i\frac{\lambda}{2\alpha}(n+h_s)=0,
 \end{eqnarray}
 with
 \begin{eqnarray}
 h_s=\frac{1}{2}+\sqrt{\frac{\omega^2}{\lambda^2}\left(\lambda-4\alpha^2\right)
 +\frac{\mu^2}{\lambda}+\frac{1}{4}}.
 \end{eqnarray}

 Quasinormal modes can be regarded as
 the resonances of a black hole under a small
 perturbation. They can be obtained
 by solving the perturbation equation in a fixed
 background geometry under the proper boundary conditions,
 which generally results in a complex damping frequency.
 The second equation of (22) can be easily solved, which
 gives the explicit expression for one family of quasinormal mode as
 \begin{eqnarray}
 \omega=\sqrt{\left(\lambda-4\alpha^2\right)\left(n+\frac{1}{2}\right)^2
 -\mu^2-\frac{\lambda}{4}}-2i\alpha\left(n+\frac{1}{2}\right)\;.
 \end{eqnarray}
 The another family of quasinormal modes is determined by the
 first equation. The solution to this equation
 is tedious, which will not be presented here.
 From Eq.(24) and the first equation of (22),
 one can conclude that the quasinormal modes
 are determined only by the parameters of black hole.

 It is noted that Eq.(22) is rather complicated to analyze.
 In order to analyze the asymptotic properties of quasinormal modes,
 we adopt a kind of identification from the calculation
 of quasinormal modes for the warped AdS$_3$ black hole
 by B. Chen and Z.-B. Xu in \cite{warpedads}.
 If making the following identification
 \begin{eqnarray}
 t\rightarrow-\varphi\;,\;\;\;
 \varphi\rightarrow\frac{t}{\alpha(r_++r_-)}\;,
 \end{eqnarray}
 one can find that the quantum
 numbers in these two backgrounds
 satisfy the following relations
 \begin{eqnarray}
 \tilde{\omega}=-\frac{m}{\alpha(r_++r_-)}\;,\;\;\;
 \tilde{m}=\omega\;.
 \end{eqnarray}
 Thus, the equation (22) can give the following
 two sectors of quasinormal modes
 \begin{eqnarray}
 &&\tilde{\omega}_L^s=-\tilde{m}
 -i\frac{\lambda(r_+-r_-)}{2\alpha(r_++r_-)}(n+\tilde{h}_s)\;,\nonumber\\
 &&\tilde{\omega}_R^s=-i\frac{\lambda}{2\alpha}(n+\tilde{h}_s),
 \end{eqnarray}
 with
\begin{eqnarray}
 \tilde{h}_s=\frac{1}{2}+\sqrt{\frac{\tilde{m}^2}{\lambda^2}\left(\lambda-4\alpha^2\right)
 +\frac{\mu^2}{\lambda}+\frac{1}{4}}.
 \end{eqnarray}
 The expressions of left sector and right
 sectors of quasinormal modes
 will be useful for area and entropy
 quantization of this black hole.
 Strictly speaking, the transition frequencies are
 changed by the identification. But we cannot find
 any other effective approach to obtain
 the transition frequency. The transition frequencies
 obtained by the identification turn out to be
 effective in the process of entropy quantization.
 We have successfully quantized the entropy and
 obtained an equally spaced spectrum
 which is generally believed to be an exact result.

 By analog to the quasinormal modes of BTZ
 black hole and warped AdS black hole,
 one can define the left and the right temperatures
 as
 \begin{eqnarray}
 &&T_L=\frac{\lambda(r_+-r_-)}{4\pi\alpha(r_++r_-)}\;,\nonumber\\
 &&T_R=\frac{\lambda}{4\pi\alpha}\;.
 \end{eqnarray}
 The relation of the left and the right temperatures
 and the Hawking temperature is then given by
 \begin{eqnarray}
 \frac{1}{T_H}=\frac{1}{T_L}+\frac{1}{T_R}\;.
 \end{eqnarray}

 In the context of AdS/CFT corresponding,
 the dual conformal field theory on the
 boundary can be separated into
 two independent left-moving and right-moving sectors
 at thermal equilibrium with different temperatures.
 The perturbation fields in the black hole
 background are dual to the operators in the boundary
 conformal field theory.
 The relaxation time $\tau$
 for a thermal state back to thermal equilibrium
 in the boundary conformal field theory
 is related to the imaginary part of
 quasinormal modes.
 The dual aspect of black hole and
 conformal field theory is very interesting
 for further study, which however is
 beyond the scope of the present work.

 In summary, we have computed the quasinormal modes
 of scalar field perturbation in this section.
 In the next section, we will
 perturb the black hole by a spinor
 field to calculate its
 corresponding quasinormal modes.

 \section{quasinormal modes of spinor field}

 In this section, we will calculate the
 quasinormal modes of fermionic field perturbation
 in the background of three dimensional G\"{o}del black hole.
 For this purpose, we should consider the spinor field $\Psi$
 with mass $\mu$ in this black hole,
 which obeys the covariant Dirac equation
 \begin{eqnarray}
 \gamma^a e^{\mu}_a\left(
 \partial_\mu+\frac{1}{2}\omega_{\mu}^{ab}\Sigma_{ab}
 \right)\Psi+\mu\Psi=0\;,
 \end{eqnarray}
 where $\omega_\mu^{ab}$ is the spin connection,
 which can be given in terms of the tetrad $e_a^\mu$,
 $\Sigma_{ab}=\frac{1}{4}[\gamma_a, \gamma_b]$, and
 $\gamma^0=i\sigma^2$, $\gamma^1=\sigma^1$,
 $\gamma^2=\sigma^3$,
 where the matrices $\sigma^k$ are the Pauli matrices.

 In order to analytically solve
 the covariant Dirac equation,
 we should firstly choose a
 proper tetrad field and calculate
 the corresponding spin connection.
 According to the metric of three dimensional G\"{o}del black hole,
 the tetrad field can be selected to be
 \begin{eqnarray}
 e^0&=&\sqrt{\Delta}d\varphi\;,\nonumber\\
 e^1&=&\frac{1}{\sqrt{\Delta}}dr\;,
 \nonumber\\
 e^2&=&dt-2\alpha r d\varphi\;.
 \end{eqnarray}
 This selection for the tetrad field
 is not unique, but simple and convenient
 in the following calculations.

 By employing the Cartan structure equation
 $de^a+\omega^a_{\;\;b}\wedge e^b=0$,
 one can calculate the spin connection directly.
 The nonvanishing components of the spin connection
 are listed as follows
 \begin{eqnarray}
 \omega_t^{01}&=&\alpha\;,\nonumber\\
 \omega_r^{02}&=&\frac{\alpha}{\sqrt{\Delta}}\;,\nonumber\\
 \omega_\varphi^{01}&=&\frac{\Delta'}{2}-2\alpha^2 r\;,\nonumber\\
 \omega_\varphi^{12}&=&\alpha\sqrt{\Delta}\;.
 \end{eqnarray}

 The inverse of the tetrad field
 is also needed
 \begin{eqnarray}
 e_0&=&\frac{2\alpha r}{\sqrt{\Delta}}\partial_t+
 \frac{1}{\Delta}\partial_\varphi\;,\nonumber\\
 e_1&=&\sqrt{\Delta}\partial_r
 \;, \nonumber\\
 e_2&=&\partial_t\;.
 \end{eqnarray}

 Assuming that the spinor field takes the form
 $\Psi=(\psi_+(r),\psi_-(r))e^{-i\omega t+im\varphi}$
 and changing the variables to $z$,
 one can finally derive the following
 equations of motion after some algebra
 \begin{eqnarray}
 z^{\frac{1}{2}}(1-z)\frac{d\psi_+}{dz}+
 \left[\left(\frac{2i\alpha r_+ \omega-im}{\lambda(r_+-r_-)}
 +\frac{1}{4}\right)z^{-\frac{1}{2}}+
 \left(-\frac{2i\alpha r_-\omega-im}{\lambda(r_+-r_-)}
 +\frac{1}{4} \right)z^{\frac{1}{2}}
 \right]\psi_+ &&\nonumber\\
 +\frac{1}{\sqrt{\lambda}}\left(
 i\omega+\frac{\alpha}{2}+\mu
 \right)\psi_-=0\;,&&\nonumber\\
 z^{\frac{1}{2}}(1-z)\frac{d\psi_-}{dz}+
 \left[\left(-\frac{2i\alpha r_+\omega-im}{\lambda(r_+-r_-)}
 +\frac{1}{4}\right)z^{-\frac{1}{2}}+
 \left(\frac{2i\alpha r_-\omega-im}{\lambda(r_+-r_-)}
 +\frac{1}{4} \right)z^{\frac{1}{2}}
 \right]\psi_- &&\nonumber\\
 +\frac{1}{\sqrt{\lambda}}\left(-i\omega+\frac{\alpha}{2}+\mu
 \right)\psi_+=0\;.&&
 \end{eqnarray}
 The above equation can also be solved by the hypergeometric
 function. The solution of $\Psi_+(z)$ with the
 ingoing boundary condition at the horizon can be explicitly
 expressed as
 \begin{eqnarray}
 \psi_+(z)=z^{\alpha_f} (1-z)^{\beta_f} F(a_f,b_f,c_f,z)\;,
 \end{eqnarray}
 with the parameters as
 \begin{eqnarray}
 &&\alpha_f=-\frac{i\left(2\alpha r_+\omega-m\right)}{\lambda(r_+-r_-)}-
 \frac{1}{4}\;,\nonumber\\
 &&\beta_f=\frac{1}{2}-\sqrt{\frac{\omega^2}{\lambda^2}
 \left(\lambda-4\alpha^2\right)+\frac{1}{\lambda}
 \left(\mu+\frac{\alpha}{2}\right)^2}\;,\nonumber\\
 &&\gamma_f=\frac{i\left(2\alpha r_-\omega-m\right)}{\lambda(r_+-r_-)}
 -\frac{1}{4}\;,\nonumber\\
 &&a_f=\alpha_f+\beta_f+\gamma_f\;,\nonumber\\
 &&b_f=\alpha_f+\beta_f-\gamma_f\;,\nonumber\\
 &&c_f=2\alpha_f+1\;.
 \end{eqnarray}
 Then, the solution of $\psi_-(z)$ can be obtained by integrating
 the second equation of (35) as
 \begin{eqnarray}
 \psi_-(z)&=&z^{-\alpha_f-\frac{1}{2}}(1-z)^{\alpha_f+\gamma_f+1}
 \int z^{c_f-1}(1-z)^{b_f-c_f-1}F(a_f,b_f,c_f,z)dz\nonumber\\
 &=&z^{\alpha_f+\frac{1}{2}}(1-z)^{\beta_f}F(a_f+1,b_f,c_f+1,z)\;.
 \end{eqnarray}

 One can also find the leading asymptotic behaviour
 $(z\rightarrow 1)$ of this solution
 \begin{eqnarray}
 \psi_+&\simeq& z^{\alpha_f}
 (1-z)^{\beta_f}\frac{\Gamma(c_f)\Gamma(c_f-a_f-b_f)}
 {\Gamma(c_f-a_f)\Gamma(c_f-b_f)}\;,\nonumber\\
 \psi_-&\simeq& z^{\alpha_f+\frac{1}{2}}
 (1-z)^{\beta_f}\frac{\Gamma(c_f+1)\Gamma(c_f-a_f-b_f)}
 {\Gamma(c_f-a_f)\Gamma(c_f-b_f)}\;.
 \end{eqnarray}
 By imposing the vanishing Dirichlet boundary
 condition at infinity, one can find the following
 relation
 \begin{eqnarray}
 c_f-a_f=-n\;,\;\;\;c_f-b_f=-n\;,
 \end{eqnarray}
 which gives two equations of quasinormal modes for
 spinor field perturbation
 \begin{eqnarray}
 &&\omega-\frac{m}{\alpha(r_++r_-)}
 +i\frac{\lambda(r_+-r_-)}{2\alpha(r_++r_-)}(n+h_f)=0,\nonumber\\
 &&\omega+i\frac{\lambda}{2\alpha}(n+h_f-\frac{1}{2})=0,
 \end{eqnarray}
 with
 \begin{eqnarray}
 h_f=\frac{1}{2}+\sqrt{\frac{\omega^2}{\lambda}
 \left(1-\frac{4\alpha^2}{\lambda}\right)+\frac{1}{\lambda}
 \left(\mu+\frac{\alpha}{2}\right)^2}.
 \end{eqnarray}

 Similar to the case of scalar field,
 the first equation is hard to solve.
 This family of quasinormal modes will
 not be explicitly given. The second equation
 can be easily solved, and the quasinormal
 modes are given by
 \begin{eqnarray}
 \omega=\sqrt{(\lambda-4\alpha^2)n^2
 -\left(\mu+\frac{\alpha}{2}\right)^2}
 -2i\alpha n\;.
 \end{eqnarray}

 One can also make the identification of (25) and (26)
 to obtain two sectors of quasinormal modes for spinor field
 similarly to scalar field
 \begin{eqnarray}
 &&\tilde{\omega}_L^f=-\tilde{m}
 -i\frac{\lambda(r_+-r_-)}{2\alpha(r_++r_-)}(n+\tilde{h}_f),\nonumber\\
 &&\tilde{\omega}_R^f=-i\frac{\lambda}{2\alpha}(n+\tilde{h}_f-\frac{1}{2}),
 \end{eqnarray}
 with
 \begin{eqnarray}
 \tilde{h}_f=\frac{1}{2}+\sqrt{\frac{\tilde{m}^2}{\lambda}
 \left(1-\frac{4\alpha^2}{\lambda}\right)+\frac{1}{\lambda}
 \left(\mu+\frac{\alpha}{2}\right)^2}.
 \end{eqnarray}

 One can see that the quasinormal modes
 of spinor field exhibit the same asymptotic
 properties as that of scalar field when $n$
 is very large. In the next section, we will
 use this observation to quantize the area and the entropy
 of three dimensional G\"{o}del black hole.

 \section{area and entropy quantization using quasinormal modes }

 The two families of the quasinormal modes (27) and (44) of the
 three dimensional G\"{o}del black
 hole for scalar field and spinor field
 at large $n$ for a fixed $|m|$, in particular for
 $n\gg |m|$, can give the asymptotic
 properties as
 \begin{eqnarray}
 &&\tilde{\omega}_L=-i\frac{\lambda(r_+-r_-)}{2\alpha(r_++r_-)}n,\nonumber\\
 &&\tilde{\omega}_R=-i\frac{\lambda}{2\alpha}n.
 \end{eqnarray}
 This asymptotic properties of the two families
 of quasinormal modes gives two possible transition
 frequencies, $\omega _{Lc}$ and $ \omega _{Rc}$.
 We find two transition frequencies
 corresponding to each quasinormal mode
 \begin{eqnarray}
 &&\omega_{Lc}=\frac{\lambda(r_+-r_-)}{2\alpha(r_++r_-)},\nonumber\\
 &&\omega_{Rc}=\frac{\lambda}{2\alpha}.
 \end{eqnarray}

 Based on Bohr's correspondence principle, it is proposed in
 \cite{cqg1,cqg2} that the transition frequency $\omega_c$ of a black hole
 in the semiclassical limit can be considered as the oscillation
 frequency in a classical system of periodic motion. Then, the
 action variable of the corresponding classical system
 with energy $E$ and transition frequency $\omega_c$ is identified
 and quantized via Bohr-Sommerfeld quantization in the semiclassical
 limit as follows
 \begin{eqnarray}
 {\mathfrak I}=\int {dE\over {\omega_c}}= n \hbar\;,
 ~~~(n=0,1,2,\cdots)\;.
 \end{eqnarray}
 Because the change of the energy $E$ of the system
 is the change of the mass $M$ of the black hole,
 one can finally obtain, in the semiclassical
 limit, the quantization condition for a black hole
 \begin{eqnarray}
 {\mathfrak I}=\int {dM\over {\omega_c}}= n \hbar\;,
 ~~~(n=0,1,2,\cdots)\;.
 \end{eqnarray}
 This formula holds for a black hole
 with quasinormal modes regardless
 of whether it is rotating or not.
 We will use this quantization condition
 to obtain the discrete area and entropy spectra
 of the three dimensional G\"{o}del black hole.

 In the present case, we should consider
 the two action variables corresponding to each possible
 transition frequency. The two quantization conditions are given
 by
 \begin{eqnarray}
 {\mathfrak I_{L}} &=& \int  { dM \over {\omega _{Lc} } }\nonumber\\
 &=&\frac{\alpha}{G\lambda}
 \sqrt{4G^2\nu^2-
 \frac{2GJ}{\alpha}\frac{(1+\alpha^2 l^2)}{l^2}}\nonumber\\
 &=&\frac{\alpha}{4G}(r_+-r_-)\nonumber\\
 &=& n_L \, \hbar ~,\\
 {\mathfrak I_{R}} &=& \int  { dM \over {\omega _{Rc} } }\nonumber\\
 &=&\frac{2\alpha}{\lambda}\nu\nonumber\\
 &=&\frac{\alpha}{4G}(r_++r_-)\nonumber\\
 &=& n_R \, \hbar ~,
 \end{eqnarray}
 where $n_L, \, n_R= 0,1,2,..$ . Notice that we have $n_R \ge n_L$.

 Firstly, we find that the total horizon area is quantized and equally
 spaced. The total horizon area is given by
 \begin{equation}
 A_{tot} \equiv A_{out}+A_ {in} = 4\pi\alpha (r_++r_-)  ~.
 \end{equation}
 So, according to the second quantization condition (51), one can
 obtain
 \begin{equation} \label{b1}
 A_{tot}=16 \,\pi G \,n_R \, \hbar  ~~,\quad ~~n_R =0,1,2,..\,.
 \end{equation}

 Next, we find that the difference of the two horizon areas
 is also quantized and equally spaced due to the first
 quantization condition. The area difference is given by
 \begin{equation}
 A_{sub} \equiv A_{out}-A_ {in} = 4\, \pi \alpha (r_+-r_-)~.
 \end{equation}
 Therefore, the area difference is quantized by
 \begin{equation}  \label{b2}
 A_{sub}=16 \,\pi G \,n_L\, \hbar   ~~,\quad ~~n_L=0,1,2,..\,.
 \end{equation}

 From (\ref{b1}) and (\ref{b2}), we further find the quantizations of the
 outer and inner horizon areas as
 \begin{equation}
 A_{out}=8\pi G  (n_R + n_L ) \,\hbar ~~,  ~~A_{in}= 8 \pi G (n_R -
 n_L) \,\hbar  ~,
 \end{equation}
 where $ n_R \ge n_L\,$.
 Therefore we find that both the outer and the inner horizon areas are
 equally spaced with the same spacing
 \begin{equation}
 \triangle A_{out}=8\pi G \hbar ~~,  ~~ \triangle A_{in}= 8\pi G
 \hbar  ~.
 \end{equation}
 Then, the entropy spectrum of three dimensional
 G\"{o}del black hole is given by
 \begin{equation} \label{t7}
 S=\frac{A_{out}}{4G}= 2 \,\pi  \,(n_R + n_L) \hbar ~.
 \end{equation}
 Hence it also has equal spacing, consistent with Bekenstein's
 proposal
 \begin{equation} \label{b_3}
 \triangle S= 2\,\pi \hbar ~.
 \end{equation}
 These results indicate that the spacing of entropy
 spectrum for three dimensional G\"{o}del black hole is consistent with that for
 BTZ black hole \cite{cqg1} and warped AdS black hole \cite{cqg2}.
 However, the quantization process from different
 viewpoints will lead to different conclusions. For example,
 when taking the real part of the quasinormal frequency
 as vibrational frequency for the non-rotating BTZ black hole,
 the equally spaced mass spectrum and the non-equally spaced
 area spectrum are obtained in \cite{setare}.

 \section{Conclusion}

 As an exact solution to Einstein-Maxwell-Chern-Simons
 theory in 2+1 dimensions with a negative cosmological constant,
 three dimensional G\"{o}del black holes are interesting
 to study because they display the same peculiar
 properties as their higher dimensional counterparts.
 In this paper, we have
 studied the perturbation of scalar field and
 spinor field in the background of three dimensional
 G\"{o}del black hole. It is shown that
 the wave equations of scalar field and spinor
 field can be exactly solved in terms of
 hypergeometric functions.
 The quasinormal modes of
 scalar field and spinor field are analytically calculated
 by imposing the Dirichlet boundary condition
 at spatial infinity,
 which are shown to be of the same form in both cases.
 By considering the physical interpretation of quasinormal modes,
 we obtain the consistent transition frequencies
 from the quasinormal modes of scalar field and spinor field.
 As an application of quasinormal modes,
 we use the proposal of area quantization of rotating
 black hole in \cite{cqg1,cqg2} to investigate area and entropy
 quantization of three dimensional
 G\"{o}del black hole. It has been discussed in Sec.II that
 there is some amount of ambiguity as to the conserved mass in
 three dimensional G\"{o}del black hole. However,
 it is shown that, when the conserved mass
 of three dimensional G\"{o}del black hole is taken as the parameter
 $\nu$, the area and entropy can be quantized and
 the spectra are equally-spaced.

 \section*{Acknowledgement}

 The author is grateful to Ming-Fan Li,
 Shi-Xiong Song, Pu-Jian Mao and Lin-Yu Jia
 for useful discussions.

 \end{document}